\magnification=\magstep1
\tolerance=500
\rightline{TAUP 2729-03}
\rightline{13 January, 2003}
\vskip 2true cm

\centerline{\bf The Conformal Metric Associated}
\centerline{\bf with the}
\centerline{\bf U(1) Gauge of the Stueckelberg-Schr\"odinger Equation}
\bigskip
\centerline{O. Oron and L.P. Horwitz \footnote{*}{Also at Department of
Physics, Bar Ilan University, Ramat Gan 52900, Israel}}
\smallskip
\centerline{Raymond and Beverly Sackler School of Physics and
Astronomy}
\centerline{Tel Aviv University, Ramat Aviv 69978, Israel}
\bigskip
\noindent {\it Abstract:\/} We review the relativistic classical
and quantum mechanics of Stueckelberg, and introduce the 
compensation fields necessary for the gauge covariance of the 
Stueckelberg-Schr\"odinger equation.  To achieve this, one must
introduce  a
fifth, Lorentz scalar, compensation field, in addition to the four
vector fields which
compensate the action of the space-time derivatives. A generalized
Lorentz force can be derived from the classical Hamilton
equations associated with this evolution function.  We show that the
fifth (scalar) field can be eliminated through the introduction of a
conformal metric on the spacetime manifold.  The geodesic equation
associated with this metric coincides with the Lorentz force, and is
therefore dynamically equivalent.  Since the generalized Maxwell
equations for the five dimensional fields provide an equation relating
the  fifth field with the spacetime density of events, one can derive
 the spacetime event density associated with the Friedmann-Robertson-Walker
solution of the Einstein equations.  The resulting density, in the
conformal coordinate space, is isotropic and homogeneous,
decreasing as the square of the Robertson-Walker scale factor.  Using
the Einstein equations, one sees that both for the static and matter
dominated models, the conformal time slice in which the events which
generate the world lines are contained becomes progressively thinner as
the inverse square of the scale factor, establishing a simple
correspondence between the configurations predicted by the underlying
Friedmann-Robertson-Walker dynamical model and the 
configurations in the conformal coordinates.
\bigskip
\noindent PACS:
\noindent keywords:conformal metric, Lorentz force, generalized Maxwell theory
\vfill
\break

\bigskip
\noindent
{\bf 1. Introduction}
\par We shall review here some of the basic foundations of a
relativistically covariant classical and quantum dynamics$^{1,2,3}$,
in which not only space coordinates and space momenta are dynamical
variables, but also the time $t$ and the energy $E$, resulting in an
8N dimensional phase space for N particles.  The dynamical evolution
of such a system is parametrized by a universal time $\tau$,
essentially the time postulated by Newton, which is
unaffected by motion or forces which may act on the system, and
provides a universal correlation between subsystems. The existence of
such a time is suggested by the early thought experiment of Einstein,
for which signals emitted at some interval $\Delta \tau$ in a frame $F$,
according to intervals of a clock in that frame, are detected in a
second inertial frame $F'$.  The time of detection, recorded in terms
of clocks (of the ``same manufacture'') in $F'$ show that the detected
interval is altered by the Lorentz time dilation.  The interval in $F'$,
measured on the same type of clocks ($\tau$) as found in the emitting frame, is
then called $\Delta t$, and participates covariantly in the Lorentz
transformation. This construction would not be possible without the
assumption that there are clocks of identical structure in both
frames, and therefore form the basis of the assumption of a universal
time.  From this argument, one sees that the rate of the clocks must
be the same.  To be able to construct a mechanics of an N body system,
in which motions of the individual parts are correlated, or to think
of an evolution of the world, it was made a fundamental assumption of
the theory that this invariant time $\tau$ is universal$^2$.
 Following Stueckelberg$^1$, consider the Hamiltonian of a free particle
to be (we take $c =1$)
$$ K = {p^\mu p_\mu \over 2M},  \eqno(1.1)$$
where $p^\mu = (E, {\bf }p)$. Since $E$ and ${\bf p}$ are assumed
independent variables, the quantity $p^\mu p_\mu = {\bf p}^2 - E^2 $
(we use the metric $(-, +, +, +)$) is not constrained to the constant
numerical value of an {\it a priori} given mass. The Hamilton
equations associated with Eq.$ (1.1)$ are
$$ {dx^\mu \over d\tau} = {\partial K \over \partial p_\mu} = {p^\mu
\over M}, \eqno(1.2)$$
and 
$$ {dp_\mu \over d\tau} = - {\partial K \over  \partial x^\mu}. \eqno(1.3)$$

Dividing the space part by the time part of the equations for the
 $\tau$ derivative of $x^\mu$,
eliminating $d\tau$, one finds that
$$ {d{\bf x} \over dt} = {{\bf p} \over E}, $$
which is the definition of velocity in special relativity.  One finds,
 following a similar procedure, all of the formulas for Lorentz
transformation as dynamical  equations follow from the Hamilton
equations generated by the free Hamiltonian $(1.1)$.   
\par The quadratic form of Eq. $(1.1)$ makes it possible to separate
variables in the two-body problem.  For example, for an
action-at-a-distance potential,
$$  K = {{p_1\,}^\mu {p_1\,}_\mu \over 2M_1} + {{p_2\,}^\mu
{p_2\,}_\mu \over 2M_2} + V(x_1  - x_2) \eqno(1.4),$$
where the potential $V$ (of dimension mass) is a scalar function of
the four-vector $x_1 - x_2$.
\par A model for the relativistic generalization of the
(spinless) Coulomb
problem was worked out classically in ref. 2, with a complete
discussion and numerical solutions in ref. 4, and quantum mechanically
in ref. 5, where quite general potentials as functions for the
invariant $(x_1 -x_2)^2$ were studied.  By the transformation
 $$ P^\mu = {p_1\,}^\mu + {p_2\,}^\mu \qquad  X^\mu = {M_1 {x_1\,}^\mu + M_2
{x_2\,}^\mu  \over M}$$        
and 
$$ p^\mu = {M_2 p_1\,^\mu - M_1 p_2\,^\mu \over M} \qquad x^\mu =
 x_1\,^\mu -x_2\,^\mu, $$
where $M= M_1 + M_2$, one obtains
$$ K = {P^\mu P_\mu \over 2M} + {p^\mu
p_\mu \over 2M_{red}} + V(x),\eqno(1.5) $$
where $M_{red} = {M_1M_2\over {M_1 +M_2}}$.  In this form, one sees
clearly that the center of mass part may be separated from the
relative motion problem. By choosing a representation for the
coordinates$^5$ (called RMS coordinates) which span the spacelike part
of the two subsectors of the spacelike region complementary to the
light cone, one finds the usual Schr\"odinger spectrum for the reduced
Hamiltonian. Taking the (constant) value of the generator $K$ at the
asymptotic ionization point to be $-M/2$ (on mass shell for both
particles), for small excitation spectrum (compared to the total mass
M), the center of mass energy spectrum was found to be $Mc^2$ plus
the correct Schr\"odinger bound state eigenvalues, plus relativistic
corrections$^5$.
\bigskip
{\bf 2. Stueckelberg-Schr\"odinger Equation}
\smallskip
\par Electromagnetism may be thought of as closely associated with the
gauge invariance of the nonrelativistic Schr\"odinger equation, in the
sense that covariance of the theory under local phase transformations
of the form $\psi \rightarrow e^{i\Lambda}\psi$ requires the addition
of compensation fields $A_i$ to the canonical momenta (acting as
 derivatives),
and the field $A_0$ to the explicit derivative $i {\partial \over
\partial t}$ on the left hand side of the Schr\"odinger equation. By
adding quadratic terms in the field strengths to the Lagrangian that
produces the Schr\"odinger equation, one finds the inhomogeneous
Maxwell equations and the equations that couple the motion of the
charged particle to the fields.  
\par In a similar way, one can demand the covariance of the
Stueckelberg theory to phase transformations (functions of spacetime
and $\tau$), and find {\it five} compensation fields, four for the
spacetime derivatives, and one for the derivative in $\tau$ generating
evolution of the system.  Defining the Hamiltonian as the function
which generates the $\tau$ derivative, i.e., putting the fifth field
(as a potential) on the right hand side, one may use the Hamilton
equations to derive a generalized Lorentz force. The Hamiltonian has
the form     
 
$$K= \eta^{\mu \nu}{(p_\mu-ea_\mu(\xi))(p_\nu-ea_\nu(\xi))
 \over 2M}-ea_5(\xi),\eqno(2.1)$$
where $\xi^\mu$ are the particle coordinates in the flat Minkowski space. 

The equations of motion using the Hamilton equations are then

$${d\xi^\sigma \over d\tau}= {\partial K \over \partial p_\sigma} = \eta^{\sigma\nu}{p_\nu-ea_\nu \over M}={p^\sigma-ea^{\sigma} \over M} \eqno(2.2)$$
$${dp_\sigma \over d\tau}= -{\partial K \over \partial \xi^\sigma}=e\eta^{\mu \nu}{\partial a_{\mu} \over \partial \xi^\sigma}{(p_{\nu}-ea_{\nu})\over M}+e{\partial a_5 \over \partial \xi^\sigma}=e{\dot \xi^{\mu}}{\partial a_{\mu} \over \partial \xi^\sigma}+e{\partial a_5 \over \partial \xi^\sigma}. \eqno(2.3) $$
We now use Eq.$(2.2)$ to substitute for ${\dot p}_\sigma$, and we finally arrive at the 5D Lorentz force:
$$ M{\ddot \xi}^{\sigma}=e{\dot \xi}^\mu f^\sigma \,_
\mu+ef^\sigma \,_5, \eqno (2.4)$$
where $f_{\alpha \beta}=\partial_{\alpha} a_{\beta}-\partial_{\beta}
 a_{\alpha}$ ($\alpha,\beta=0,1,2,3,5$); $x^5=\tau$.

We now suggest replacing the 5-potential in a flat space picture by a
new Hamiltonian which contains only a 4-potential, and takes into
account the fifth potential in the metric of a curved space picture .
We shall designate the curved coordinates with $\hat x$. 
The generator of motion is 
$$ K_r=g^{\mu \nu}{(p_\mu-ea_\mu({\hat x}))(p_\nu-ea_\nu({\hat x})) \over 2M} \eqno(2.5)$$ 
Assuming that this functional of $(p_{\sigma},{\hat x}^{\sigma})$
gives Hamilton equations which are equivalent dynamically to those of
the flat space we find:
$$ { d{\hat x}^{\sigma} \over d\tau}={\partial K_r \over \partial p_\sigma}=g^{\sigma \nu}{d \xi_{\nu} \over d\tau} \eqno(2.6)$$
This equation gives us the transformation law $d {\hat
x}^\sigma=g^{\sigma \nu}{d \xi_{\nu}}$ and  $d{\xi}_\sigma=g_{\sigma
\mu}d{ \hat x}^{\mu}$.  The second Hamilton equation gives
$${dp_\sigma \over d\tau}=-{\partial K_r \over \partial {{\hat x}^\sigma}}=-{M \over 2}{\partial g^{\mu \nu} \over \partial {{\hat x}^\sigma}} {\dot \xi}_{\mu}{\dot \xi}_{\nu}+g^{\mu \nu}{\partial a_{\mu} \over \partial  {\hat x}^\sigma}{(p_{\nu}-ea_{\nu})\over M} \eqno(2.7) $$
We now replace $ {\hat x}$ with $\xi$ using the transformation law and substitute Eq.($3$) for ${\dot p}_{\sigma}$:
$$e{\dot \xi^{\mu}}{\partial a_{\mu} \over \partial \xi^\sigma}+e{\partial a_5 \over \partial \xi^\sigma}=-{M \over 2}{\partial g^{\mu \nu} \over \partial {{\xi}_\alpha}}g_{\alpha \sigma} {\dot \xi}_{\mu}{\dot \xi}_{\nu}+g^{\mu \nu}e{\partial a_{\mu} \over \partial  {\xi}_\alpha}g_{\alpha \sigma}{\dot \xi}_{\nu} \eqno(2.8) $$

The functionals $K, K_r$ are different; however, on the physical trajectories
they take the same value, $\bf{K}={M \over 2}$, where {\bf K} is the common
numerical value.
We now show that choosing a conformal metric $g^{\mu \nu}={\Phi}({\hat x})\eta^{\mu \nu}$, the Lorentz force derived from $K_r$ is the same as the one derived from $K$.
In this case we have:
$$ g_{\mu \nu}={1 \over \Phi({\hat x})}\eta_{\mu \nu} \,\,\,\,
{\Phi}({\hat x})={1 \over 1+{e\over \bf{K}} a_5({\hat x})}. $$
In this case Eq.($2.8$) gives
$$ e{\partial a_5 \over \partial \xi^\sigma}=-{M \over 2}{1 \over \Phi}{ \partial \Phi\over \partial \xi_{\sigma}} \eta^{\mu \nu} {\dot \xi}_{\mu}{\dot \xi}_{\nu}. \eqno(2.9) $$
Using ${M \over 2}\eta^{\mu \nu}{\dot \xi}_\mu{\dot \xi}_\nu={\bf{K}
\over \Phi}$ we find that Eq.($2.9$) is indeed satisfied. This shows that the the Hamilton equations for the two generators are identical.

It is now interesting to examine the geodesic motion of this dynamical
system, assuming the fields are static in $\tau$. 
The Lagrangian in this case is 
$$L=p_{\mu}{\dot x}^\mu-H= {M \over 2}g_{\mu \nu}{\dot x}^\mu{\dot
x}^\nu+{\dot x}^\mu a_\mu. \eqno(2.10) $$
We now make a small variation in $x^\mu$
$$ x^\mu \rightarrow x^\mu+\delta x^\mu$$
$$ \delta S =\int d\tau \bigl[ {M \over 2} \bigl ({\partial g_{\mu \nu} \over \partial x^\sigma}{\dot x}^\mu{\dot x}^\nu \delta x^\sigma+2g_{\mu \nu}{\dot x}^\mu{d \delta x^\nu \over d\tau} \bigr)+ea_\mu{d \delta x^\mu \over d\tau}+{\partial a_\mu \over \partial x^\sigma}{\dot x}^\mu{\delta x}^\sigma \bigl]$$
From the minimal action principal we obtain, by  integration by parts
of the $\tau$ derivatives ($\tau$ independence of the field implies ${d \over d\tau}={\dot x}^\mu {\partial \over \partial x^\mu}$)
$$ 0= {1 \over 2} \bigl ({\partial g_{\mu \nu} \over \partial x^\sigma}{\dot x}^\mu{\dot x}^\nu-2g_{\sigma \nu}{\dot x}^\mu{\dot x}^\sigma-2g_{\sigma \nu}{\ddot x}^\nu \bigr)+{e\over M}\bigl(-{\dot x^\mu}{\partial a_\sigma\over \partial x^\mu}+{\partial a_\mu \over \partial x^\sigma}{\dot x}^\mu \bigl);$$
multiplying by $g^{\lambda \sigma}$ we finally get
$$ {\ddot x}^\lambda=-{\Gamma}^{\lambda}_{\mu \nu}{\dot x}^\mu{\dot
x}^\nu+{e \over M}{\dot x}^\mu f^{\lambda}_{\, \mu} \eqno(2.11)$$
where $f^{\lambda}_{\,\mu}=g^{\lambda \sigma}f_{\sigma \mu} $ and
$f_{\sigma \mu}={\partial a_\mu \over \partial x^\sigma}-{\partial
a_\sigma\over \partial x^\mu}$.
\bigskip
\noindent
{\bf 3. The Friedmann-Robertson-Walker Universe}
\smallskip
\par In the ``flat space'' Robertson-Walker model$^6$ (for the spatial geometry characterized by k=0) the metric
$$ds^2= d\tau^2-\Phi^2(\tau)\bigl(dx^2+dy^2+dz^2\bigr). \eqno(3.1)$$
 can be brought to the form
$$ds^2=\Phi^2(t)\bigl( dt^2-dx^2-dy^2-dz^2\bigr). \eqno(3.2)$$
 by using the transformation
$$t=\int{d\tau \over \Phi(\tau)}; \eqno(3.3)$$
$\tau$ is the time coordinate of a freely-falling object and therefore
coincides with our notion of universal $\tau$. The function
$\Phi(\tau)$ is often designated by $R$ or $a$ and is the
(dimensionless) spatial scale of the expanding universe.
In the conformal coordinates the time-coordinate is therefore related 
to $\tau$, according to the transformation above, through
$${dt \over d\tau}= {1 \over \Phi}. \eqno(3.4)$$
 It is interesting to use the Lorentz force in order to achieve the
same result. Let us assume that $a_5$ depends on $t$ alone.  In this
case, the force is
$$ {\ddot t}={e \over M}f^0_{\,5}=-{e \over M}{da^5 \over dt} \eqno(3.5)$$
The relation
$$ \Phi^2={1 \over 1+{e \over \bf{K}}a^5} \eqno(3.6)$$
then implies
$$ 2 {d\Phi \over dt}\Phi=-\Phi^4{e \over \bf{K}}{da^5 \over dt},$$
i.e.,
$${da^5 \over dt}={ {\bf K}\over e}{d \over dt}\bigl({1 \over \Phi^2}\bigr)$$
We substitute this in the force equation and multiply by $2 {\dot t}$
to obtain
$$ {d {\dot t}^2\over d\tau}=-2{{\bf K}\over M}{d \over d\tau}\bigl({1
\over \Phi^2}\bigr). \eqno(3.7)$$
Finally,  putting $\bf{K}=-{M \over 2}$ we arrive at the remarkable result
$${dt \over d\tau}={1 \over \Phi},$$
which coincides with the transformation $(3.4)$ from the time on the
freely falling clock $\tau$ to the redshifted $t$ in the conformal
form of the Robertson-Walker metric.  We see that this $t$ corresponds
to the Einstein time satisfying the dynamical Hamilton equations, and
the conformal factor of the Robertson-Walker metric coincides with the
conformal facter of the curved space embedding.
\par In this construction we have assumed the $a_5$ field to depend on
$t$ alone.  The generalized Maxwell equations then provide a simple
connection between the Robertson-Walker scale and the event
density.
\par The generalized Maxwell equations$^3$ are
$$ \partial_\alpha f^{\beta \alpha} = j^\beta, \eqno(3.8)$$
where $j^\beta = (j^\mu, \rho)$ satisfies $\partial_\beta j^\beta =
\partial_\mu j^\mu + \partial_5 \rho = 0$, and $\rho$ is the event
density.  In the generalized Lorentz gauge $\partial_\alpha a ^\alpha
=0$, we have
$$ -\partial_\alpha \partial^\alpha a^5 = j^5 = \rho. \eqno(3.9)$$
Since $a^5$ depends on t alone $(3.9)$ becomes
$$ \partial_t^2  a^5 = \rho. \eqno(3.10)$$
From $(3.6)$, 
$$ a_5={{\bf K} \over e}\bigl({1 \over \Phi^2}-1\bigr)$$
so that from $(3.10)$
$$\rho=-{2 {\bf K} \over e} \bigl[ {\Phi_{tt} \over \Phi^3}-3{\Phi_t^2 \over \Phi^4} \bigr] \eqno (3.11)$$
\par  The space-time geometry is related to the density of matter
$\rho_M$ through the Einstein equations
$$ G^{\mu \nu}\equiv R^{\mu \nu}-{1 \over 2}g^{\mu \nu}R=8 \pi G
T^{\mu \nu}, \eqno(3.12) $$ 
where $R^{\mu \nu}$ is the Ricci tensor, $R$ is the scalar curvature
and $T^{\mu \nu}$ is the energy-momentum tensor. For the perfect fluid
model (isotropy implies the $T^{\mu \nu}$ is diagonal)
$$T^{\mu \nu}=\rho u^\mu u^\nu+P(g^{\mu \nu}+u^\mu u^\nu). \eqno(3.13)$$
The $(0,0)$ component (referring to $\tau$) is then 
$$T^{\tau \tau}=8 \pi G \rho_M. \eqno(3.14)$$
using the affine connection derived from the metric $(3.1)$ one finds
$$ G^{\tau\tau}= 3 {{\dot{\Phi}}^2 \over \Phi^2}=8 \pi G \rho_M \eqno(3.15)$$
and the (equal) diagonal space-space components are (for example, we
write the $x,x$ component
$$G^{xx} = - {1 \over \Phi^2}\bigl[2 {{\ddot a}\over a}+{{\dot a}^2
\over a^2}\bigr]=8 \pi G T^{xx} = {8 \pi G P \over
\Phi^2}. \eqno(3.16)$$
\par Since $T^{\tau\tau}$ is the $(0,0)$ component of a tensor, it
follows from $(3.14)$ and $(3.4)$ that the matter density in the conformal
coordinates is given by
$$ \rho'_M = { 1 \over \Phi^2} \rho_M. \eqno(3.17)$$
To establish a connection between the density of events in spacetime
$\rho$ and the density of matter(particles) $\rho'_M$, in space at a
given time $t$, we assume that,
$$ \rho'_M=\rho \Delta t \eqno(3.18)$$
where $\Delta t$ is the time interval (in the conformal coordinates associated with the Stueckelberg evolution) in which the events generating the particle world lines are uniformly spread. 
It then follows from $(3.18)$ that 
$$  \Delta t = {\rho_M \over \rho \Phi^2}. \eqno(3.19)$$
\par We now consider two examples.  For the static universe, for $\rho_M$
constant, it follows from Eq. $(3.15)$ that $\Phi$ is given by an
exponential; it then follows from $(3.11)$ that $\rho$ is constant, so
that 
$$ \Delta t \propto \Phi^{-2}. \eqno(3.20)$$
\par For the matter dominated universe, where the pressure is
negligible$^6$, one sees from $(3.16)$ that 
$$ 2\Phi \Phi_{tt}= \Phi_t^2, $$
and substituting in $(3.11)$, one finds after changing $\tau$
derivatives to $t$ derivatives in $(3.15)$ that ${\rho_M \over \rho}$ is
constant.  It then follows that $\Delta t \propto \Phi^{-2}$ in this
case as well. 
\par This result implies that, at any given stage of development of
the universe, i.e., for a given $\tau$, the events generating the
world lines lie in an interval of the conformal time $t$ which becomes
smaller as $\Phi$ becomes large in the order of $\Phi^{-2}$ With the
relation $(3.4)$, this corresponds, on the other hand, to a narrowing
distribution, of order $\Phi^{-1}$ in $\tau$, contributing to a set of
events observed at a given value of the conformal time
$t$.   In general, if one observes the configuration of a system at a
given $t$, the events detected may have their origin at widely
different values of the world time $\tau$ parametrizing the
trajectories (world lines) of the spacetime events.  It would be
generally difficult to relate such configurations to the
configurations in spacetime (at a given $\tau$, instead of at a given
$t$) predicted by a dynamical theory.  However, in this case, we see
that the spreading is narrowed for large $\Phi$, so that the set of
events occurring at a given $\tau$ is essentially the same as the set
of particles occurring at a given $t$.  The observed configuratoins
therefore become very close to those predicted by the underlying
dynamical model. In the general case, the relation between $\rho
(\tau)$ and $\rho_M(t)$ could be very complicated, and it may be
difficult to see in the observed configurations a simple relation to
the dynamical model evolving according to the world time.
 In the static and matter dominated Friedmann-Robertson-Walker model, the
correspondence between the dynamical theory and observed
configurations becomes more clear as $\Phi$ becomes large.
 
\par 
\bigskip
\noindent
{\bf 4. Summary and Conclusions}
\smallskip
\par We have shown that the fifth potential of the generalized Maxwell
theory, obtained throught the requirement of gauge invariance of the
Stueckelberg-Schr\"odinger equation, can be eliminated in the function
generating evolution of the classical system by replacing the
Minkowski metric in the kinetic term by a conformal metric.  The
Hamilton equations resulting from this function coincide with the
geodesic associated with this metric, and with the Hamilton equations
of the original form, i.e., the geodesic equations of the conformal
metric describe orbits that coincide with solutions of the original
Hamilton equations, as found in previous work which studied the
replacement of an invariant (action-at-a-distance) potential by a
conformal metric$^7$.  In this case the geodesic equations are those
obtained from the conformal geodesic with the addition of a Lorentz
force in standard form.
\par The Robertson-Walker metric can be put into conformal form. The
conformal factor of the Robertson-Walker metric can then be put into
correspondence with the $a^5$ field of the generalized Maxwell theory
and therefore, through its $t$ derivatives (we assume no explicit
$\tau$ dependence) with the event density. In both the static and the
matter dominated models, the set of events generating the world
lines of the expanding universe condense into progressively thinner
slices of the conformal time. This result implies that the
observed configuration of the universe at a given conformal time $t$,
for large $\Phi$ approximately corresponds  to the configuration in spacetime
predicted by the Friedmann-Robertson-Walker model at a given $\tau$. 

\bigskip
\noindent
{\it References}
\smallskip
\frenchspacing
\item{1.}  E.C.G. Stueckelberg, Helv. Phys. Acta {\bf 14}, 322
(1941); {\bf 14}, 588 (1941).
\item{2.} L.P. Horwitz and C. Piron,
 Helv. Phys. Acta {\bf 46}, 316 (1973).
\item{3.}  M.C. Land, N. Shnerb and L.P. Horwitz,
Jour. Math. Phys. {\bf 36}, 3263 (1995); N. Shnerb and L.P. Horwitz,
Phys. Rev. {\bf 48A}, 4068 (1993).
\item{4.}M.A. Trump and W.C. Schieve, {\it Classical Relativistic
Many-Body Dynamics,} Kluwer Academic, Dordrecht (1999).
\item{5.} R. Arshansky and L.P. Horwitz, Jour. Math. Phys. {\bf 30},
66,380 (1989).
\item{6.} S. Weinberg, {\it Gravitation and Cosmology}, Wiley, New
York (1972); R.M. Wald, {\it General Relativity}, University of Chicago
Press, Chicago (1984).
\item{7.}D. Zerzion, L.P. Horwitz and R. Arshansky,
Jour. Math. Phys. {\bf 32}, 1788(1991).
 
\vfill
\end
\bye